\documentclass[iop]{emulateapj}
\usepackage{graphicx}
\usepackage{amssymb}
\usepackage{url}

\shorttitle{Live fast, die young}
\shortauthors{R. Gobat et al.}

\begin{document}

\title{The early early type: discovery of a passive galaxy at $\lowercase{z_{spec}}\sim3$}

\author{R. Gobat\altaffilmark{1}, V. Strazzullo\altaffilmark{1}, E. Daddi\altaffilmark{1}, 
M. Onodera\altaffilmark{2}, A. Renzini\altaffilmark{3}, M. B\'{e}thermin\altaffilmark{1}, 
M. Dickinson\altaffilmark{4}, M. Carollo\altaffilmark{2}, A. Cimatti\altaffilmark{5}}

\altaffiltext{1}{Laboratoire AIM-Paris-Saclay, CEA/DSM-CNRS--Universit\'e Paris Diderot, 
Irfu/Service d'Astrophysique, CEA Saclay,Orme des Merisiers, F-91191 Gif sur Yvette, France}
\altaffiltext{2}{Institute for Astronomy, ETH Z\"{u}rich, Wolfgang-Pauli-strasse 27, 8093 Z\"{u}rich, 
Switzerland}
\altaffiltext{3}{INAF - Osservatorio Astronomico di Padova, Vicolo dell'Osservatorio 5, I-35122 Padova, 
Italy}
\altaffiltext{4}{National Optical Astronomy Observatory, P.O. Box 26732, Tucson, AZ 85726, USA}
\altaffiltext{5}{Dipartimento di Astronomia, Universit\`{a} di Bologna, Via Ranzani 1, I-40127 Bologna, 
Italy}

\begin{abstract}
We present the discovery of a massive, quiescent galaxy at $z=2.99$. We have obtained a \emph{HST}/WFC3 spectrum of 
this object and measured its redshift from the detection of a deep 4000 \AA\ break consistent with an old population 
and a high metallicity. By stellar population modeling of both its grism spectrum and broad-band photometry, 
we derive an age of $\sim0.7$~Gyr, implying a formation redshift of $z>4$, and a mass $>10^{11}$~$M_{\sun}$. 
Although this passive galaxy is the most distant confirmed so far, we find that it is slightly less compact than 
other $z>2$ early-types of similar mass, being overall more analogous to those $z\sim1.6$ field early-type galaxies. 
The discovery of this object shows that early-type galaxies are detectable to at least $z=3$ and suggests that the 
diversity of structural properties found in $z=1.4-2$ ellipticals to earlier epochs could have its origin in a variety 
of formation histories among their progenitors.
\end{abstract}


\section{\label{intro}Introduction}

In the nearby Universe, most stars reside in massive, passively evolving spheroidal systems, either spiral 
bulges or elliptical galaxies \citep{Bal04}. Early-type galaxies (ETG), such as ellipticals and lenticulars, 
are often the most massive galaxies in their surrounding environment and ETGs include the dominant stellar
component in clusters of galaxies. As far as they can be found in significant numbers, massive ETGs exhibit 
very homogeneous spectrophotometric properties, requiring that their stellar content formed rapidly at relatively 
high redshifts and evolved passively for most of the cosmic  time. In particular, studies of the core ETG
population in galaxy clusters suggest that the quenching of their star formation (SF) was already underway by $z\sim3$.

The precursors of $z\lesssim1$ massive elliptical galaxies can arguably be found in some high-redshift 
submillimeter-selected galaxies, whose extreme star-formation rates 
\citep[SFRs, in excess of $1000$~M$_{\sun}$~yr$^{-1}$;][]{Dad09,Mic10} imply that they should exhaust their gas 
reservoirs on short time scales \citep[e.g.,][]{Car10} and thus offer a striking example of rapid, unsustainable 
growth at early epochs. The spheroidal morphology of ETGs might then be either the consequence of merging events, 
triggering the starburst or happening after the quenching of SF, or dynamical instabilities internal 
to the galaxies \citep[e.g.,][]{Mar09,Bou09}. This latter process may actually be the dominant one, given that 
the vast majority ($\sim 86\%$) of ETGs are fast rotators \citep{Ems11}.\\

One would therefore expect to find quenched, massive ETGs already at $z\sim2-3$ and indeed passively evolving 
galaxies (PEGs; generally color selected as passive and not necessarily morphologically early-type) have been spectroscopically 
confirmed up to $z\sim2.5$ \citep{vanD08}. Most of these early passive systems appear to be up to a factor $\sim10$ denser 
than their modern counterparts \citep[e.g.,][]{Dad05,Ci08,vanD08}. However, the search for high-redshift ETGs is hampered not 
only by the expected rarity of such objects, but also by  the intrinsic difficulty in reliably detecting them and measuring 
their redshifts. The most distinctive spectral feature of ETGs, the pronounced metal break at 4000 \AA\ is redshifted 
to the near-infrared at $z>1$ and detection, let alone characterization, of $z\sim2$ ETGs requires good continuum 
signal to noise (S/N) only achievable with extremely long exposures from the ground. Consequently, the sample 
of confirmed $z\gtrsim2$ ETGs is still small, and its high-redshift tail, i.e., the earliest epoch at which passive 
galaxies can be found, poorly constrained.\\

\begin{figure*}
\centering
\includegraphics[width=0.23\textwidth]{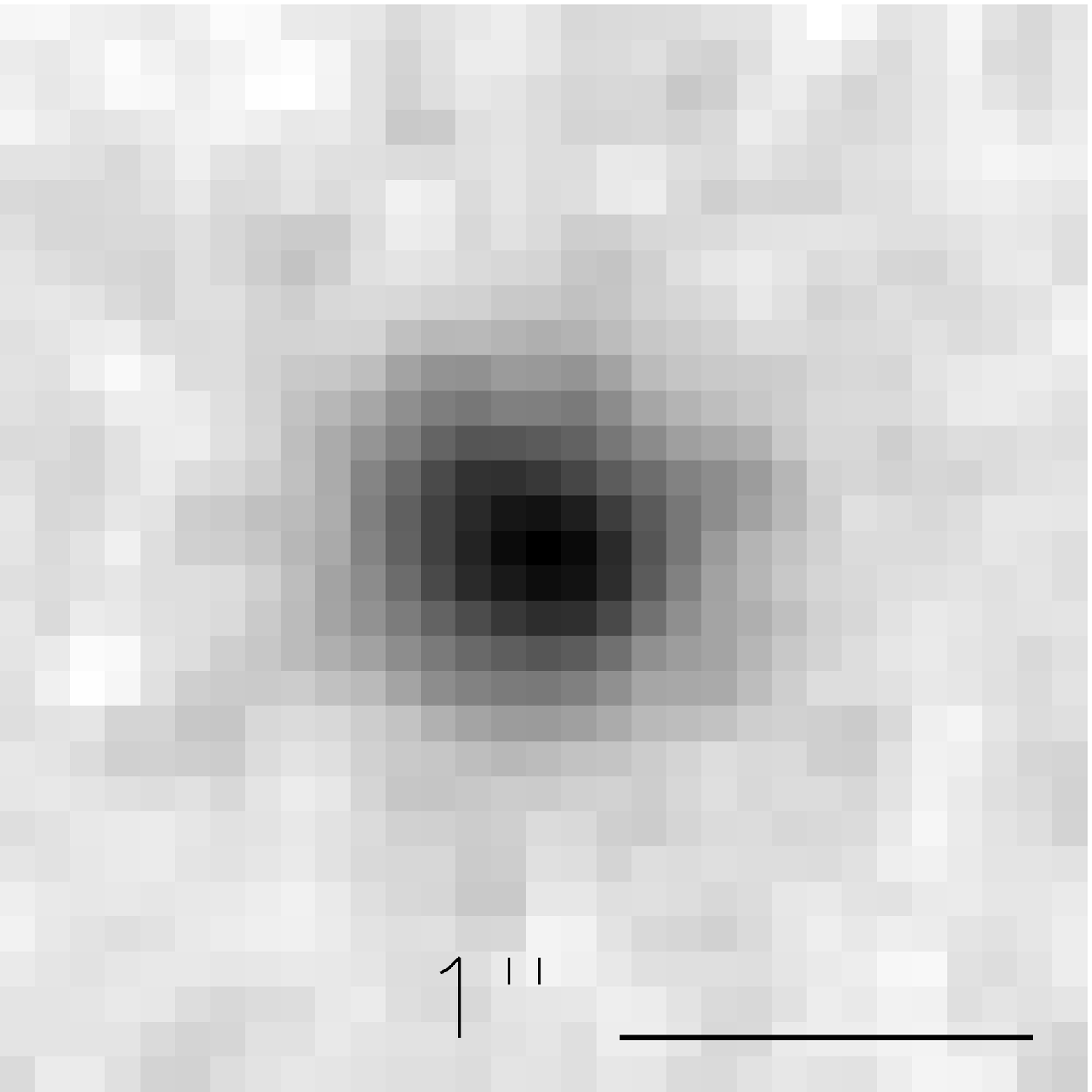}
\includegraphics[width=0.64\textwidth]{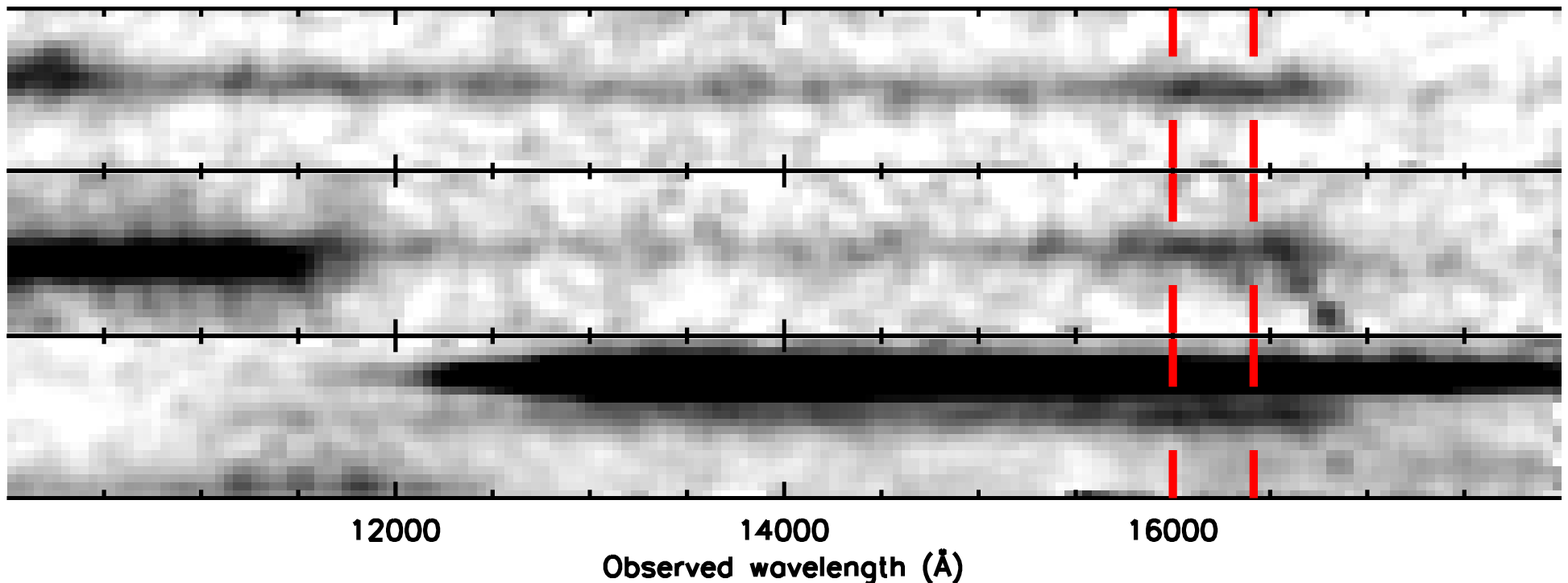}
\caption{Left: $2\arcsec.6\times2\arcsec.6$ \emph{HST}/WFC3 $F140W$ image of RS-235. The image is 2\arcsec.6$\times$2\arcsec.6
wide, with a pixel size of 0\arcsec.06. Right: two-dimensional spectra of RS-235 in each of the three orientations, 
between 1 and 1.8~$\mu$m, smoothed with a 2~pixel bin. The bottom spectrum suffers from slight contamination between 1.25 
and 1.5~$\mu$m from the tail of a bright, parallel second order trace and the two others have overlapping traces below 
1.2~$\mu$m. Each pixel is 46.5~\AA\ ($\sim11.7$~\AA\ in rest-frame) wide in the dispersion direction and $\sim$0\arcsec.13 
in the spatial (vertical) direction. The red lines bracketing the spectrum show the position of the 4000~\AA\ break and 
H$_{\delta}$, respectively.}
\label{fig:2d}
\end{figure*}

Here we present the serendipitous discovery of the most distant spectroscopic ETG so far identified, i.e., a massive and 
passive galaxy at $z\sim3$. Throughout the Letter, we assume a $\Lambda$CDM cosmology with $H=70$~km~s$^{-1}$~Mpc$^{-1}$, 
$\Omega_m=0.27$ and $\Lambda=0.73$. Magnitudes are given in the AB photometric system.

\section{\label{obs}Observations}

This object (hereafter RS-235) was discovered in deep \emph{HST}/WFC3 observations of the $z\sim 2$ cluster Cl J1449+0856 
\citep{Gob11} at 14:49:16.5, +8:55:34.7, about 56\arcsec\ from the center of the structure. These data cover a 
6.4~arcmin$^2$ field centered on Cl J1449+0856 and consist of 2 orbits of imaging with the $F140W$ filter and 16 
orbits of spectroscopy with the G141 grism, in three different orientations. The individual frames were sky-subtracted 
and reduced using the best available calibration files. Standard procedures were applied: the WFC3 images were combined 
with MultiDrizzle and grism spectra for each orientation extracted with the aXe pipeline \citep{Kum09}.
\footnote{See e.g., http://www.stsci.edu/hst/wfc3/analysis/grism\_obs/\allowbreak{}cookbook.html} 
The $F140W$ image of RS-235, as well as the two-dimensional spectra in the three orientations, is shown in Figure 
\ref{fig:2d}.Trace contamination was modeled using the $F140W$ deep image ($m_{140}=26.5$~AB at 5$\sigma$ in a 1\arcsec\ 
aperture) and the multiwavelength dataset already available for this field. This extensive set of ground-based observations 
was first described in \citet{Gob11} and has recently been extended with deep $U$ and $V$-band VLT/FORS2 imaging and deeper 
3.5-4.5~\micron\ \emph{Spitzer}/IRAC imaging, in addition to the \emph{HST}/WFC3 data. At longer wavelengths, 24~\micron\ 
\emph{Spitzer}/MIPS data were already available and \emph{Herschel}/PACS 100 and 160~\micron\ observations have recently been 
obtained to expand the dataset (the maps reach a 5$\sigma$ depth of 1.5 and 2.8~mJy respectively and will be described in a 
forthcoming paper). We have used this new multiwavelength ($U$-band to IRAC) catalog to compute photometric redshifts for all 
objects in the \emph{HST}/WFC3 field with EAZY using the standard set of templates \citep{Bram08,Bram11}.\\

RS-235 stood out in this dataset as having a 2D spectrum with a pronounced break at $\sim1.6~\mu$m, a well-constrained 
photometric redshift of $z\sim3$ and rest-frame $UVJ$ colors consistent with an old stellar population 
\citep{Wuy07}. 
Furthermore, RS-235 is not seen at 24, 100 or 160~\micron\ although, due to its redshift and the depth of the maps, 
this non-detection only provides very loose constraints on its SFR, with an upper limit of SFR$<400$~$M_{\sun}$~yr$^{-1}$ 
assuming recent templates from \citet{Mag12}.

To ensure that the spectrum of RS-235 is free of contamination so that its continuum shape can be measured reliably,
we performed, for each of the three WFC3 orientations, two different extractions from the cutout produced by aXe. We first 
used a 3~pixel aperture, roughly corresponding to the FWHM of the object, and linearly fitted the observed spectrum with 
the combined contamination and trace models. We compared the resulting one-dimensional spectrum with that obtained from 
fitting each cross-dispersion column of the cutout with profiles derived, for each trace, from the profile of the 
corresponding object in the $F140W$ image.
The decontaminated spectra were then stacked to produce a single high S/N ($\sim11$ pixel$^{-1}$ at 1.6 $\mu$m) spectrum 
and converted to flux units.
We note that, while the first orientation suffers from moderate contamination between 1.25 and 1.5 $\mu$m (from the tail 
of a bright non overlapping trace) and the other two below 1.25~$\mu$m (from overlapping traces), the region redward of 
1.5 $\mu$m where the break is seen is entirely free from contamination in all three cases. The sampling of the spectrum 
is 46.5~\AA\ with the G141 grism and its FWHM resolution 126~\AA, as grism spectra are convolved with the brightness profile 
of the dispersed sources. In this case, the profile is sufficiently circular ($b/a\sim0.8$) that it was not necessary to 
optimize the orientation of the pseudo-slit during extraction. 

\begin{figure*}
\centering
\includegraphics[width=0.49\textwidth]{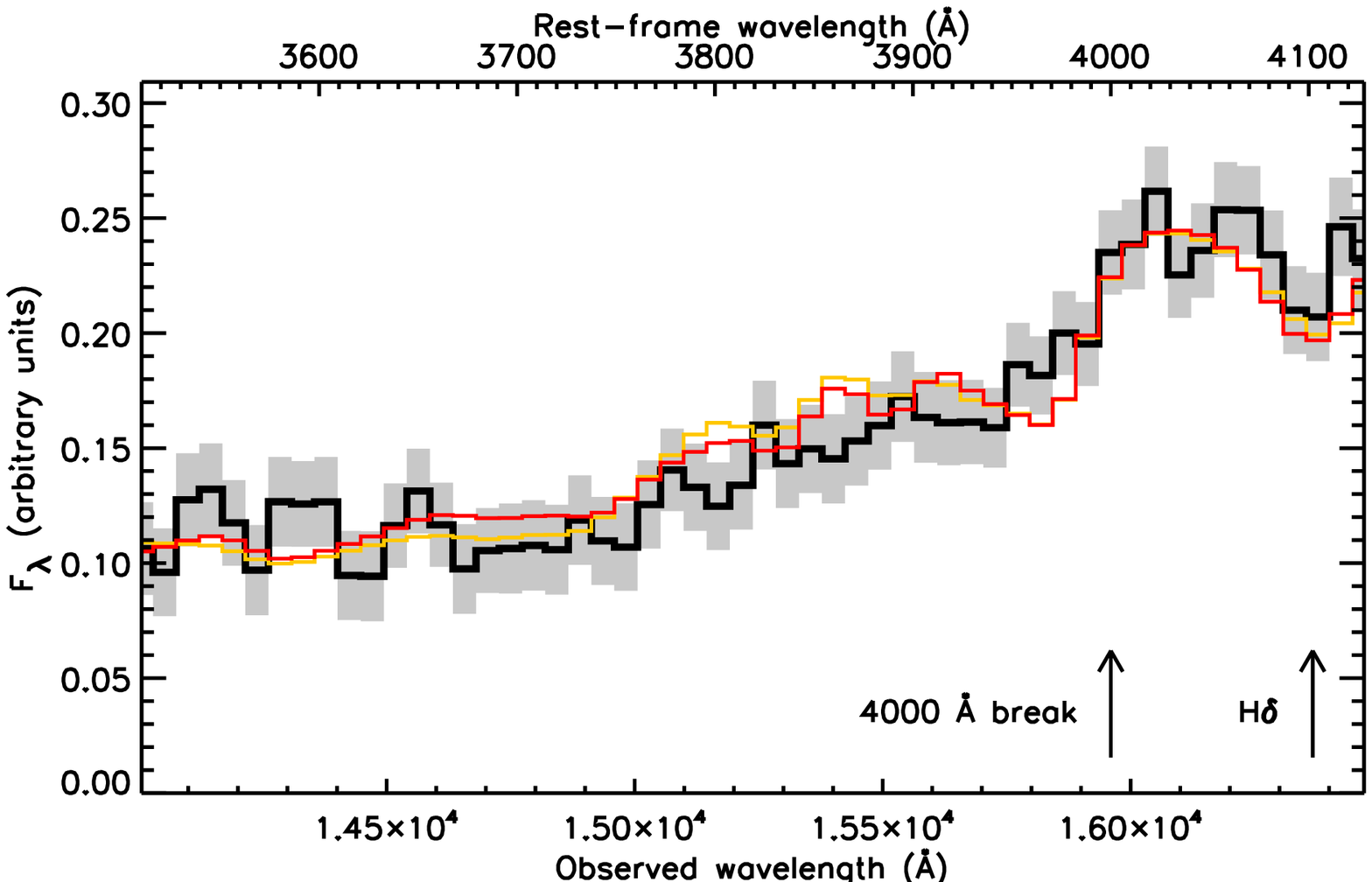}
\includegraphics[width=0.49\textwidth]{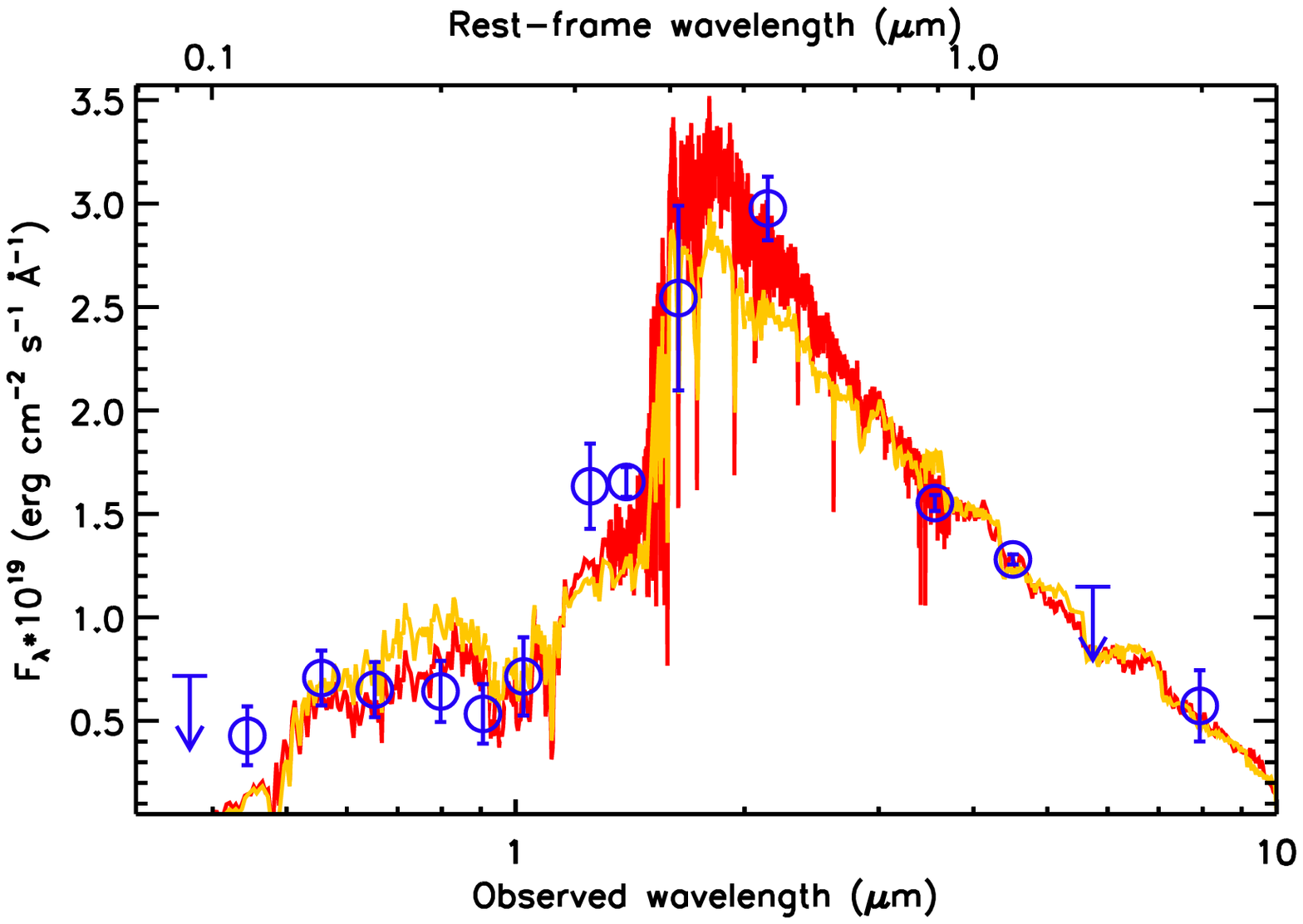}
\caption{Left: \emph{HST}/WFC3 G141 spectrum of RS-235 (black) and associated errors (gray), after subtraction of the 
contamination and stacking of the three orientations, compared with the best solution to the combined fit to the SED and 
spectrum (BC03 in red, M05 in orange). Right: 15-band ($UBVRIzYJm_{140}HK_s$+IRAC) SED of RS-235 (blue) and associated 
errors (3$\sigma$ upper limits are shown as arrows) with the same BC03 and M05 best solutions (red and orange, respectively) 
to the combined fit.}
\label{fig:sedfit}
\end{figure*}

\section{\label{results}Analysis}

As shown in Figure \ref{fig:sedfit}, the stacked spectrum shows two distinct spectral features: the break at 1.6~$\mu$m 
and an absorption feature (H$_{\delta}$) at 1.636~$\mu$m. Their position suggests a redshift of $z\simeq 3$, which 
was measured by comparison with templates (see below), yielding $z=2.993\pm0.015$. As its spectrophotometry precludes dusty 
SF solutions (see below), this makes RS-235 the most distant PEG confirmed so far.

\subsection{\label{sedfit}Spectrophotometric modeling}

To estimate the stellar mass and characterize the stellar population of the galaxy, we compared both the combined 15-band 
spectral energy distribution (SED) and high S/N part (1.4--1.65~$\mu$m) of the stacked grism spectrum to a range of 
stellar population models. These were built using high resolution \citet{BC03} and \citet{M05} (hereafter BC03 and M05) single 
stellar population templates, assuming $Z=(1-2)Z_{\sun}$ (see below), a \citet{Salp55} initial mass function (IMF). 
We considered two different sets of star formation histories (SFH): a delayed, exponentially declining SFH with characteristic 
time scale $\tau$ between 1~Myr and 2~Gyr and ages from 100~Myr to the age of the Universe at $z=3$, and a rising exponential 
SFH of the form SFR$(t)=M_0\times A\times e^{At}$, starting at $z_0=10$. The first one can be considered a versatile, albeit 
simplistic, ``generic'' SFH, while the second scenario is meant to put lower limits to the average SFR and formation time scales 
required to produce a galaxy such as RS-235. 
It is appropriate for  a galaxy on the ``main sequence'' (MS) of star forming galaxies and relies on the apparent 
weak evolution of the specific SFR (sSFR) at $z\gtrsim2.5$ \citep[e.g.,][]{Mag10,Gon10,Kar11}, which indeed implies an 
exponential growth. Its two parameters are the initial mass $M_0$ of the galaxy at the beginning of MS evolution,\ and the 
parameter $A$ setting the time scale and determined by the normalization of the SFR-$M^{\star}$ relation at $z\sim2.5$. A 
shallower slope of the SFR-$M^{\star}$ relation, e.g., a slope $\sim 0.8$ found by \citet{Rod11}, would imply a power-law SFH 
with SFR$(t)\propto t^4$, whereas the exponential form has a simpler dependence on $M_0$. We allowed values up to 0.7~dex below 
and above the MS.

In both cases we allowed for a SF cut-off at an arbitrary time  to be determined by the best fit procedure, with an option 
for a secondary burst after the initial quenching contributing up to 50\% of the final mass. This addition is motivated by 
the detection of the galaxy in the rest-frame far-UV (the $V$ band, as shown in Figure \ref{fig:sedfit}) implying the presence 
of a small amount of massive stars. Note that the detection in $B$ is marginal, at $\gtrsim3\sigma$, and might be due to 
contamination from a nearby source.  
Dust extinction was also added using a \citet{Cal00} extinction law and recombination lines were computed using Gaussian 
profiles of matching resolution, assuming the SFR-$H\alpha$ calibration of \citet{Ken98} and standard line ratios \citep{And03}. 
The model spectra were convolved with the filters responses to produce synthetic SEDs, smoothed to the resolution of the grism 
spectrum using a Gaussian kernel and rebinned to match the observed spectrum. The results of the fit are shown in Figure 
\ref{fig:sedfit} and in Table \ref{tab:results}.\\

Due to the low resolution of the grism spectrum, the spectral fit is mainly constrained by the shape of the continuum, 
namely the break at $\sim$1.6 $\mu$m and the blueward slope of the spectrum, and the lack of prominent spectral features. 
Notably, the spectrum does not show any [OII]3727 emission. We find a mean stellar age of 
$t_{SF}=0.7^{+0.15}_{-0.1}$~Gyr (corresponding to $z_f>4$), with a last episode of SF contributing less than 10\% 
of the final mass of $M_{\star}=1.2^{+0.5}_{-0.2}\times10^{11}$~$M_{\sun}$. For the delayed exponential SFH, the fit yields 
time scales of $\leq200$~Myr. In the exponential SFH case, solutions with sSFR below the fiducial MS value ($\sim3$~Gyr$^{-1}$ 
at $10^{11}$~$M_{\sun}$, using \citet{Bet12}) require apparently unrealistic seed masses of $>10^{9}$~$M_{\sun}$ at 
$z=10$ (see e.g., M. Sargent et al., in preparation). In this case, the fit implies doubling time-scales of 220~Myr or less.
The amount of extinction compatible with the SED is small, $E(B-V)<0.1$, and the fit accordingly rules out star-forming 
solutions, yielding a negligible residual SFR of $0-0.5$~$M_{\sun}$~yr$^{-1}$. RS-235 thus appears to have sSFR of less than 
$5\times10^{-3}$~Gyr$^{-1}$, at least 2.7~dex below the MS of star forming galaxies at $z\sim3$. 
Interestingly, models based on \citet{M05} templates provide a less good fit to the rest-frame optical-NIR SED, underpredicting 
the flux in the observed $K_s$ band (rest-frame 5000--6000~\AA). As RS-235 is still a young stellar population, this could 
be due to, e.g., the effect of thermally pulsing-asymptotic giant branch stars in the M05 models (which would produce a 
shallower Rayleigh-Jeans slope) or the presence of convective overshooting in the stellar tracks used by the BC03 models 
\citep{M06}.\\

\begin{deluxetable*}{ccccccc}
\tablecaption{Physical and Structural Parameters from Modeling\label{tab:results}}
\tablehead{\colhead{$z$} & \colhead{D$_{\textrm{n}}$4000} & \colhead{log$M_{\star}$} & \colhead{$t_0$\tablenotemark{a}} & 
\colhead{$t_{SF}$\tablenotemark{b}} & \colhead{$r_e$} & \colhead{$n$}\\
\colhead{} & \colhead{} & \colhead{($M_{\sun}$)} & \colhead{(Gyr)} & \colhead{(Gyr)} & \colhead{(\arcsec)} & \colhead{}}
\startdata
$2.993\pm0.015$ & $1.62^{+0.09}_{-0.08}$ & $11.08^{+0.15}_{-0.12}$ & $0.9^{+0.8}_{-0.2}$ & $0.7^{+0.15}_{-0.10}$ & 
$0.16^{+0.02}_{-0.02}$ & $1.8\pm0.3$\\
\enddata
\tablenotetext{a}{Look-back time to the beginning of star formation}
\tablenotetext{b}{SF-weighted age, i.e., look-back time at which the galaxy formed half of its stars}
\end{deluxetable*}

\subsubsection{Metallicity}

\begin{figure}
\centering
\includegraphics[width=0.45\textwidth]{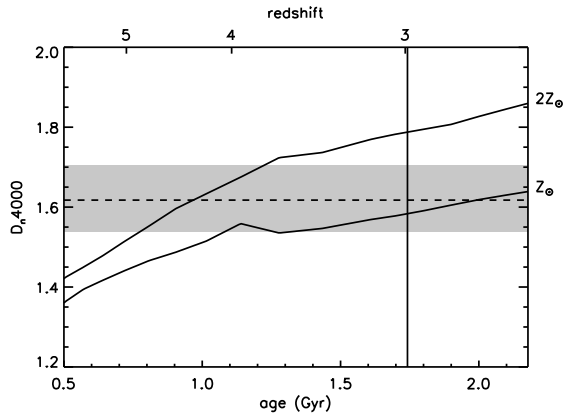}
\caption{D$_{\textrm{n}}$4000 index as a function of age and redshift, for unreddened SSPs of solar and twice solar metallicity 
formed at $z=10$. The dashed horizontal line and gray area show the value of D$_{\textrm{n}}$4000, and associated uncertainty, 
of RS-235 and the vertical line marks its redshift.}
\label{fig:met}
\end{figure}

Interestingly, solar metallicity models are rejected at $>3\sigma$ by the stellar population modeling. Quantifying 
the strength of the 4000~\AA\ break using the D$_{\textrm{n}}$4000 index \citep{Balo99}, we find 
D$_{\textrm{n}}4000=1.62\pm0.08$ (where uncertainties were estimated through Monte Carlo simulations based on the error 
spectrum). As shown in  Figure \ref{fig:met}, this value is comparable to that of a maximally old ($z_f=10$ SSP) 
$Z=Z_{\sun}$ stellar population, implying that this is a lower limit to the metallicity of RS-235 and suggesting that the 
strength of the break requires not only a passive stellar population but also a stellar metallicity above the solar value.
We note that the significantly worse resolution of the \emph{HST}/WFC3 G141 spectra, compared to that of the slit spectra of lower 
redshift objects, on which the D$_{\textrm{n}}$4000 index is usually measured, can affect the apparent strength of the break. 
While this effect depends on the stellar population mix and is most pronounced in the case of SSPs, we find that at our 
effective resolution the correction is negligible for the best fitting template to the spectrum.

\subsection{\label{morph}Structural parameters}

\begin{figure*}
\centering
\includegraphics[width=0.8\textwidth]{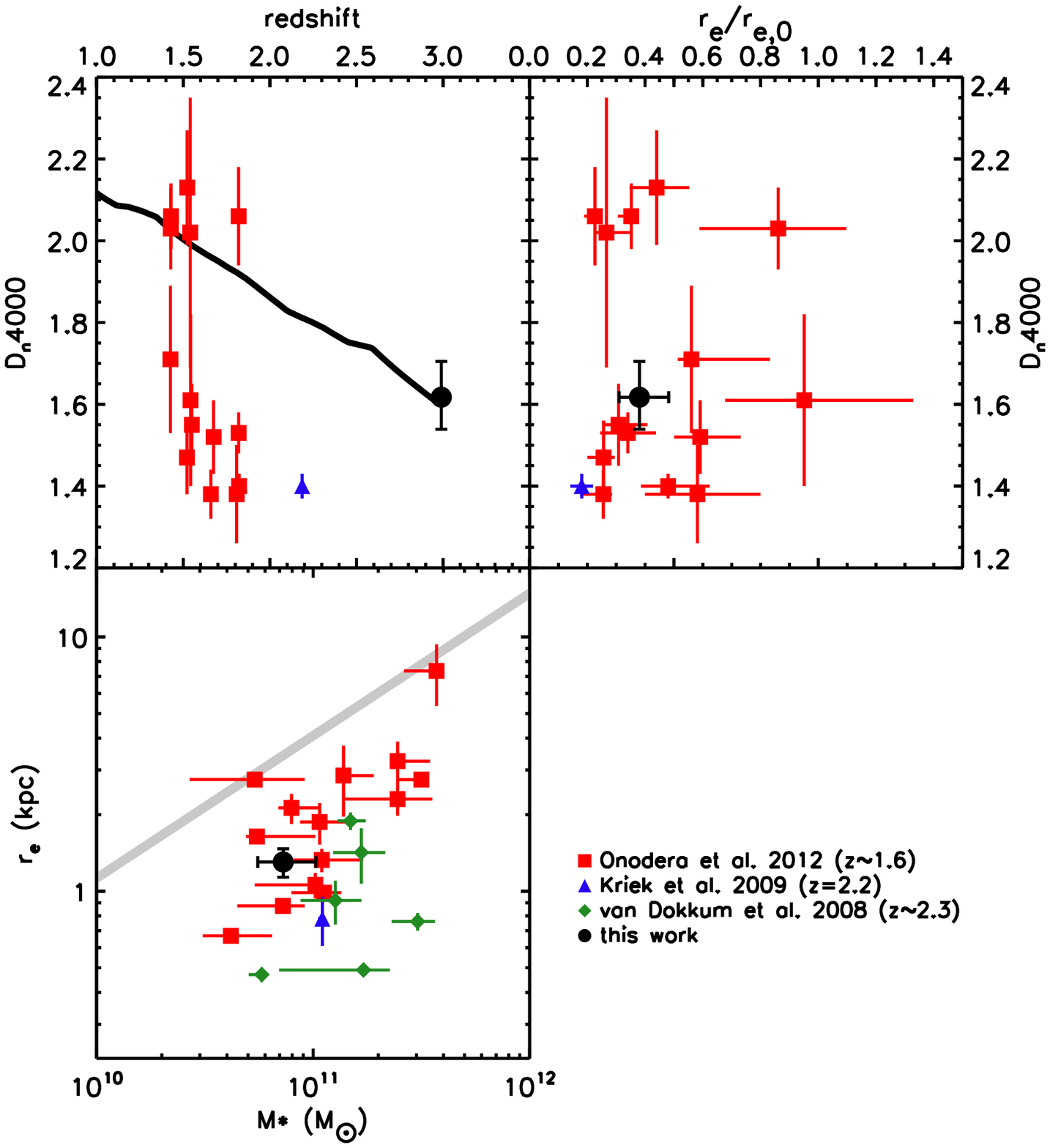}
\caption{Comparison of spectral and structural properties of RS-235 and other high redshift field PEGs 
\citep{Ono12,Kri09,vanD08}, excluding objects with disk-like ($n\lesssim1$) morphology. Top, left: D$_{\textrm{n}}$4000 
as a function of redshift. The black line shows the evolution of D$_{\textrm{n}}$4000 of RS-235, if the galaxy were to 
evolve passively to lower redshifts. 
Top, right: D$_{\textrm{n}}$4000 vs. the circularized effective radius as a fraction of the value expected from the 
local mass-morphology relation \citep{Shen03}. Bottom, left: circularized effective radius as a function of the stellar 
mass (converted to a \citet{Chab03} IMF), with the local relation shown as a gray line.}
\label{fig:d4k}
\end{figure*}

RS-235 is a resolved object in the $F140W$ image, with a FWHM size of $\sim2.5$ resolution elements ($\sim$0\arcsec.32). 
We measured its morphological properties by fitting the surface brightness profile with a S\'{e}rsic model 
\citep{Ser63} using GALFIT 3 \citep{Peng10}. We constructed a median point-spread function by combining six unsaturated, high 
S/N stars in our \emph{HST}/WFC3 field. RS-235 is relatively isolated, with the nearest sources being at least 2\arcsec.3 away 
and 1.5~mag fainter. The morphological analysis is therefore stable whether we include the closest neighbors or not. 
From a series of GALFIT simulations we find that the bias on structural parameters is negligible for a 24~mag object like this 
galaxy. The scatter on $r_e$ and $n$ is about 10\% and 20\% respectively, which we adopt as the standard error on the parameters. 
Accordingly, the S\'{e}rsic fit yields a circularized effective radius of $r_e=0.16\pm$0\arcsec.02, or $1.3$~kpc at $z=2.99$, and 
a S\'{e}rsic index of $n=1.8\pm0.3$. In Figure \ref{fig:d4k} we compare structural and stellar population parameters of RS-235 with 
samples of PEGs in the field at $z\sim1.6$ \citep[]{Ono12} and $z\sim2.3$ \citep[]{vanD08,Kri09}.\\

We note that we can not \emph{a priori} discount the possibility that RS-235 be lensed by the cluster Cl J1449+0856, 
as it is relatively close to it, and its observable properties (mass and size) therefore affected. To estimate this effect, 
we assumed a simple lens model with a single halo of mass $10^{14}$~$M_{\sun}$ (conservatively twice the X-ray mass, to 
account for the fact that the structure is likely not entirely virialized) at $z\sim2$, centered on the X-ray emission 
\citep[see][]{Gob11}. Considering both singular isothermal or \citet{NFW96} profiles with $c\geq1$, we derive a magnification 
of less than 3\%, almost a factor 10 lower than the uncertainties on either mass or size and thus negligible. We similarly 
estimate that the closest projected neighbors of RS-235 do not generate any appreciable lensing effect. On the other hand, we 
find a magnification of up to 7\% for a relatively bright spiral galaxy at 5\arcsec.2 and $z_{phot}=0.9$, implying 
that the stellar mass could be overestimated by this much and the size by $\sim3$\%. Although these factors are lower than the 
systematic error on either parameter, we added 7\% and 3\% to the lower uncertainties on the mass and circularized 
effective radius, respectively, to take the effect of magnification into account.

\section{\label{discuss}Discussion}

RS-235 appears to be a quenched, if not entirely passive galaxy at $z=2.99$, the most distant spectroscopically confirmed 
so far. Its spectrophotometric properties suggest that this galaxy formed the bulk of its stars relatively quickly at $z\gtrsim4$ 
and already has a high metallicity, comparable to that of local ETGs \citep[e.g.,][]{Gal06}. 
With $n\sim1.8$, it has a flatter profile than the typical passive population at lower redshifts and appears to have not acquired 
yet a characteristic ETG morphology. As with the vast majority of $z>1.5$ PEGs, it is also more compact 
\citep[2--3 times below the local mass-size relation for ETGs;][]{Shen03} than present-day ETGs, although apparently 
less so than other $z>2$ passive galaxies of similar mass and rather well within the variation of $z\sim1.6$ galaxies shown in Figure 
\ref{fig:d4k}. We note that our size measurement was made on a rest-frame near-UV (350~nm) image, as were the measurements by 
\citet[]{Ono12}, while \citet[]{vanD08} used rest-frame optical (480~nm) data. Based on the stellar population modeling, we estimate 
that the flux contribution of young ($<500$~Myr) stars increases only by 10\%--30\% from 480 to 350~nm, thus, this younger component, 
if present, should not affect much the effective radius measurement. As, additionally, the morphology of high-redshift ETGs has not been 
observed to vary significantly between the rest-frame near-UV and optical \cite[e.g.,][]{Cas10}, it is unlikely that the somewhat 
lower surface density of RS-235 compared to the \citet[]{vanD08} sample could be due to a positive age gradient. On the other hand, 
the population of PEGs at $z=1.4-2$ has been found to be diverse in terms of structural properties with some early-types having a size 
comparable to that of local ETGs \citep{Sar11,Ono10,Man10}. The existence of a $z=3$ passive galaxy, very close to its formation epoch, 
as extended as the $z\sim1.6$ population reinforces this picture and suggests that the diversity of $z\lesssim2$ ETGs could reflect not 
only a spread in assembly histories but also the existence of different formation pathways for these early ETGs.\\

For example, the compactness of high-redshift ETGs is expected to depend on whether they were produced through gas-rich or 
gas-poor mergers \citep[e.g.,][]{Spr05,Bou11}, or through secular gas exhaustion. However, the available data do not allow us to 
favor a particular formation scenario, although the relatively low S\'{e}rsic index is not indicative of a recent (major) merger.
On the other hand, the spectrophotometric fit tends to favor solutions with short doubling time scales and accelerated 
evolution at high redshift, with respect to our fiducial MS.\\

Finally, as shown in Figure \ref{fig:d4k}, there are five non-disky PEGs in the \citet[]{Ono12} sample with D$_{\textrm{n}}$4000 
equal or higher than that of RS-235 at the same epoch, assuming passive evolution of the best fit models. These galaxies are likely 
to be at least as old as RS-235 and to have been already quenched  at $z\sim3$. Their surface density, $0.08\pm0.04$~arcmin$^{-2}$, 
is compatible with our single detection in a 6.4~arcmin$^2$ field, suggesting that they and RS-235 come from the same early PEG 
population. It also implies that confirmation of further $z\sim3$ PEGs would require relatively few appropriately deep WFC3 
pointings and should thus be quite feasible, although the response of the G141 grism will limit the identification of the 
4000~\AA\ break to $z<3.1$ until newer instruments, such as NIRSpec on the \emph{James Webb Space Telescope}, become available.

\acknowledgements
RG, VS, ED and MB were supported by grants ERC-StG UPGAL 240039 and ANR-08-JCJC-0008.

\end{document}